# TOP QUARK PHYSICS AT THE DØ EXPERIMENT

A.P. HEINSON

*Department of Physics, University of California,
Riverside, CA 92521–0413, USA*

for the DØ Collaboration

ABSTRACT

In this paper I present the latest results on top quark physics from the DØ collaboration since the discovery of the top quark in March 1995. I summarize the discovery results, discuss progress since the discovery, and show how we can measure the top quark mass using three separate techniques. The measurements were made at the Fermilab Tevatron, a $p\bar{p}$ collider with $\sqrt{s}$ = 1.8 TeV, using ~50 pb$^{-1}$ data collected from 1992 to early 1995.

## 1. Introduction

Prior to the discovery of the top quark earlier this year, a surprising amount was known about its properties. A direct search for $t\bar{t}$ pair production by the CDF collaboration in 1992 using 4.1 pb$^{-1}$ of data[1] led to a 95% confidence level (CL) lower limit on the top quark mass of 91 GeV. This limit was extended by the DØ collaboration using 13.5 pb$^{-1}$ of data[2] in 1994 to give a 95% CL lower limit of 131 GeV. In mid-1994, CDF announced evidence for $t\bar{t}$ production from 19.3 pb$^{-1}$ data[3], with 12 events found when only 5.7 ± 0.5 were expected from known background processes. Since some of these events were double-$b$-tagged, this formed a 2.8 $\sigma$ effect. In DØ, we reanalyzed our 13.5 pb$^{-1}$ of data in light of our 131 GeV lower mass limit, to search for higher mass top pairs[4]. We found 9 events with a background of 3.8 ± 0.9 events, a 1.9 $\sigma$ effect consistent with no top production. These results were all in agreement with the indirect evidence for the existence of the top quark from fits to precision electroweak data from LEP, SLD and neutrino scattering experiments[5]. The fits at that time gave a mass window for the top quark of 150 to 210 GeV, with the range determined principally by the choice of Higgs boson mass. So the stage was set for what was to come when DØ and CDF analyzed their much larger data sets from the Tevatron Run 1b with ~50 pb$^{-1}$ integrated luminosity.

## 2. Top Quark Pair Production at the Tevatron

The Tevatron Collider at Fermi National Accelerator Laboratory near Chicago in the United States is a proton-antiproton machine operating at a center of mass energy of $\sqrt{s}$ = 1.8 TeV. At this energy, top quarks are produced predominantly in pairs, with ~90% of the cross section coming from $q\bar{q} \to t\bar{t}$ interactions and the remaining ~10% from gluon fusion $gg \to t\bar{t}$. In the analysis that follows, we assume that the top quark decays 100% of the time as prescribed by the Standard Model (SM), i.e. $t \to W^+b$ and $\bar{t} \to W^-\bar{b}$. This is a good approximation, since the CKM matrix element $V_{tb}$ which links the top and

---

bottom quarks is very close to unity: $0.9988 \leq |V_{tb}| \leq 0.9995$ at the 90% CL[6], assuming only three quark generations. We classify the various decay channels of the $t\bar{t}$ pairs by how the *W* bosons decay. When both *W*'s decay to an electron or muon (either directly or via a tau decay $W \rightarrow \tau\nu_\tau$, $\tau \rightarrow e\nu_e$ or $\mu\nu_\mu$), then the $t\bar{t}$ channel is called *dilepton* production. When one *W* boson decays leptonically as above, and the other decays hadronically ($W^+ \rightarrow u\bar{d}$ or $c\bar{s}$ and charge conjugate decays) then the mode is known as *lepton+jets* production. Finally, when both *W*'s decay hadronically, then this forms the *alljets* channel. Decay modes with a hadronically decaying tau lepton in the final state are not considered separately in this analysis, but form a small subset of the lepton+jets modes from $\tau e$, $\tau\mu$ and $\tau\tau$ production, and a similar small subset of the alljets channel from $\tau$+jets and $\tau\tau$ production. The branching fractions for the dilepton decay modes are ~1.2% each for *ee* and $\mu\mu$ decays, ~2.5% for the $e\mu$ mode, ~15% for each of the lepton+jets modes (*e*+jets and $\mu$+jets), and ~45% for the alljets channel, plus small contributions from the tau channels.

The analysis strategy adopted by DØ for searching for $t\bar{t}$ events was to select events with isolated leptons, jets and missing transverse energy (evidence for one or more neutrinos in the event) and to determine the event selection efficiencies from data. We measured all the principal backgrounds and many of the minor ones directly from the data, and checked the background calculations using a looser event selection which allowed many more events to pass the cuts. Relative to the most recently published DØ results[5], the search which led to the discovery had a data set approximately four times as large, the new analysis reduced the backgrounds by a factor of between three and six (depending on the decay channel) and the signal acceptance was kept between 70 and 80% of the previous value.

### 3. The DØ Detector and Particle Identification

The DØ detector is described in detail elsewhere[7]. For this analysis, the critical elements included a nonmagnetic central tracking system for tracking electrons, muons and jets and rejecting photons, a calorimeter for measuring the transverse energy of leptons and jets and the total missing transverse energy for inferring the passage of a neutrino, and an outer muon spectrometer for measuring the momentum of the muons. Particle identification algorithms are also described in detail elsewhere[8]. Since there is no central magnetic field, DØ cannot distinguish between electrons and positrons and they will be known generically as "electrons" here. An electron is identified by the presence of an electromagnetic cluster of energy in the calorimeter with a track pointing towards it from the primary vertex. The energy cluster must be isolated from other clusters, and in the central pseudorapidity region of the detector. It must not be identified as a pion by the transition radiation detector. A muon is identified by a central track which connects the primary vertex to a track passing through the muon spectrometer, via a track through the calorimeter along which is deposited the energy from approximately one minimum ionizing particle. If the muon is considered to have come from a *W* boson decay, then it must be isolated from any jets in the calorimeter and from any other leptons in the event. If it is considered to have come from the semileptonic decay of a *B* hadron in a jet (from the *b* quark from the top decay) then it must be close to or within the radius of a jet. It is then known as a tagging lepton, or "$\mu$ tag" and is used to identify a jet as a "*b* jet".

## 4. Data Search

(a) *Dilepton Channels*

The principal advantage of search for $p\bar{p} \to t\bar{t}X \to ee$, $\mu\mu$, $e\mu + X$ is that these dilepton channels are very clean, i.e. there are almost no background processes that mimic the signals. The disadvantage is that the branching fractions are very low. The cross section for $t\bar{t}$ dilepton events is only $3.5 \times 10^{-12}$ of the total $p\bar{p}$ cross section, for a top quark mass of 180 GeV. The main backgrounds are dilepton decays of Z bosons with accompanying initial state radiation to fake the b jets from top, Drell Yan production, $W^+W^-$ +jets, WZ, multijets with $b\bar{b}$ or $c\bar{c}$ decaying semileptonically, and instrumental fakes where a jet is misidentified as an electron and there is significant mismeasured missing transverse energy. Table 1 shows the cuts applied in the dilepton search for each channel. $H_T$ is the scalar sum of the transverse energies of the jets and the electron (if present) in the event. To study the backgrounds more carefully, the $H_T$ cut is removed. The cuts are then known as a *loose* selection. Table 2 shows the results of the search for top pair production in the dilepton channels.

| Cut | $e\mu$ | $ee$ | $\mu\mu$ |
|---|---|---|---|
| Lepton $p_T$ [GeV] | > 15 / 12 | > 20 / 20 | > 15 / 15 |
| Missing $E_T$ [GeV] | > 20 | > 25 | > 0 |
| Number of Jets | ≥ 2 | ≥ 2 | ≥ 2 |
| Jet $E_T$ [GeV] | > 15 | > 15 | > 15 |
| $H_T$ [GeV] | > 120 | > 120 | > 100 |
| not a Z | | | Z kinematic fit |

Table 1  Dilepton event selection cuts.

| Channel | $e\mu$ | $ee$ | $\mu\mu$ | Total |
|---|---|---|---|---|
| $\int Ldt$ [pb$^{-1}$] | 47.9 ± 5.7 | 55.7 ± 6.7 | 44.2 ± 5.3 | |
| MC $t\bar{t}$ Signal | 0.34 ± 0.04 | 0.25 ± 0.05 | 0.11 ± 0.02 | 0.70 ± 0.07 |
| Background | 0.12 ± 0.03 | 0.28 ± 0.14 | 0.25 ± 0.04 | 0.65 ± 0.15 |
| Data | 2 | 0 | 1 | 3 |

Table 2  Dilepton search results. The $t\bar{t}$ MC calculation is for a top quark of mass 200 GeV, using the ISAJET generator.

From approximately 50 pb$^{-1}$ of data, we found two $e\mu$ events and one $\mu\mu$ event, with an expected background of 0.65 ± 0.15 events. The ISAJET[9] Monte Carlo signal prediction for the number of events expected is for a top quark mass of 200 GeV. Figure 1 shows the distribution in $H_T$ of the three dilepton channels of $t\bar{t}$ production, and the respective backgrounds. It can be seen that the variable $H_T$ provides good discriminating power between signal and background. $H_T$ may be considered to be a measure of the temperature of the event.

(b) *Lepton+Jets Channels*

The main advantage of using the lepton+jets channels to search for $t\bar{t}$ production is that the branching fraction is significantly higher than for the dilepton channels. This advantage is counterweighed by the much higher backgrounds. The backgrounds to these channels are principally $W$+jets events and multijets where a jet fakes an electron and there is significant mismeasured missing transverse energy.

We have used two independent analysis techniques to separate $t\bar{t} \rightarrow e+$ jets and $\mu+$ jets from the background processes. We split the data set into two nonoverlapping sets, one set containing a tagging muon which was used to identify a $b$ jet, and the other set without such a muon, to which we applied a topological selection procedure. The variables used for event shape selection included $H_T$, here defined as the scalar sum of the transverse energy of all the jets, and aplanarity[10] $A$, a measure of how planar the event is. The tagging muon had to be nonisolated, i.e. close to or inside a jet, with the opposite definition to that used to label a muon as isolated (one from a $W$ boson decay). Since requiring a tagging muon is quite a stringent demand, we relax the number of jets required from 4 to 3 to regain acceptance, remove the aplanarity cut completely and loosen up the $H_T$ cut. Table 3 shows the event selection cuts used for lepton+jets events. The loose cuts involved removing the $H_T$ cut and relaxing the aplanarity cut from 0.05 to 0.03 in the nontag event sample. Table 4 shows the results of these cuts on the search for $t\bar{t}$ pair production in the lepton+jets channels.

| Cut | $e$+jets | $\mu$+jets | $e$+jets/$\mu$ | $\mu$+jets/$\mu$ |
|---|---|---|---|---|
| Lepton $p_T$ [GeV] | > 20 | > 15 | > 20 | > 15 |
| Missing $E_T$ [GeV] | > 25 | > 20 | > 25 | > 20 |
| Number of Jets | $\geq 4$ | $\geq 4$ | $\geq 3$ | $\geq 3$ |
| Jet $E_T$ [GeV] | > 15 | > 15 | > 20 | > 20 |
| Aplanarity ($A$) | > 0.05 | > 0.05 | > 0 | > 0 |
| $H_T$ [GeV] | > 200 | > 200 | > 140 | > 140 |
| Tagging $\mu$ [GeV] | none | none | > 4 | > 4 |

Table 3   Lepton+jets event selection cuts.

| Channel | $e$+jets | $\mu$+jets | $e$+jets/$\mu$ | $\mu$+jets/$\mu$ | Total |
|---|---|---|---|---|---|
| $\int Ldt$ [pb$^{-1}$] | 47.9 ± 5.7 | 44.2 ± 5.3 | 47.9 ± 5.7 | 44.2 ± 5.3 | |
| MC $t\bar{t}$ Signal | 1.84 ± 0.31 | 0.95 ± 0.24 | 0.81 ± 0.16 | 0.41 ± 0.10 | 4.01 ± 0.44 |
| Background | 1.22 ± 0.42 | 0.71 ± 0.28 | 0.85 ± 0.14 | 0.36 ± 0.08 | 3.14 ± 0.53 |
| Data | 5 | 3 | 3 | 3 | 14 |

Table 4   Lepton+jets search results. The $t\bar{t}$ MC calculation is for a top quark of mass 200 GeV.

Background subtraction for the lepton+jets channels was done in the following manner. Two independent methods were used. The first invoked *Berends Scaling*[11], which notes that the ratio of the number of events with at least $n$ jets to the number of events with at least $n-1$ jets should be a constant for any number of jets. Therefore, the numbers of events with 1 or more jets and with 2 or more jets (after subtraction of the multijet background where there is no real *W* boson) are used to predict the numbers of events with 3 or more jets and with 4 or more jets. The multijet background is subtracted first, because although it too should obey Berends scaling, the ratio between events with different numbers of jets is slightly different than that in the *W*+jets background. This technique was cross-checked with an event density fit in the ($A$–$H_T$) plane and the results were found to be consistent with each other. After all cuts were applied to the *l*+jets events with no $\mu$ tag, there remained 8 candidate events with a background of $1.9 \pm 0.5$ events.

The idea behind looking for a muon in or near a jet to label it as a *b* jet is as follows. Every $t\bar{t}$ event has two *b* jets from the direct decays of the top quark and $\bar{t}$ antiquark. The *b* quark can decay semileptonically via $b \rightarrow c\mu\nu_\mu$ to produce a final state muon and also the daughter charm quark can decay via $c \rightarrow s\mu\nu_\mu$ to give a second muon. Either or both of these oppositely charged muons may be detected and used to label the *b* jet. Using the branching fractions for these semileptonic decays, one can see that ~44% of all $t\bar{t}$ events will have at least one muon in them associated with a *b* jet. DØ's probability for reconstructing such a muon in or near a jet is ~45%, leading to an overall result that ~20% of all $t\bar{t}$ events will have an observed muon tag. The corresponding probability to tag a random background jet (the *mistag probability*) is ~0.4%.

In the analysis, the tagging probability was measured as a function of the number of jets in the event, the transverse energy of those jets and the amount of missing transverse energy. This probability was used to calculate the number of expected background events in the *l*+jets/$\mu$ channels. Backgrounds in these channels come mainly from *W*+jets events and multijet events where at least one jet is from a *b* or *c* quark which has decayed semileptonically to give a muon. The probability for a particle in a jet to punchthrough into the muon spectrometer to fake a muon in the jet is negligible. The measured probability for tagging a jet with a muon was convoluted with the spectrum of *W*+jets events where there is no tag to give the total predicted background from *W*+jets and multijet events with a $\mu$ tag. This technique was cross-checked using the dijet and $\gamma$+jets data samples and found to give consistent results. After all the cuts had been applied to the *l*+jets/$\mu$ events, there were 6 candidate events left, with a background of $1.2 \pm 0.2$ events. Figure 2 shows the number of events versus the inclusive jet multiplicity for the data after subtraction of the multijet events (i.e. nominally just *W*+jets events), for the *W*+jets events which have no $\mu$ tag convoluted with the measured tag rate, and the *W*+jets events with an appropriate amount of Monte Carlo $t\bar{t}$ production added. It can be seen that the data show a significant excess over the untagged *W*+jets events convoluted with the tag probability, which is compensated for by the addition of the Monte Carlo $t\bar{t}$ events. Also, the excess between data and Standard Model expectation without $t\bar{t}$ production increases as the number of jets goes up, in line with expectations for the high multiplicity top events.

(c) *Data Search Summary*

The results of the DØ search for $t\bar{t}$ production in the seven separate decay channels are shown in Table 5 for both the standard set of cuts and for the loose set. The loose cuts were designed to let more background events into the sample to enable us to check our background calculations. It can be seen that the calculated number of background events does indeed increase in a similar manner to the number of events which pass the loose selection cuts. We calculate the probability that the background fluctuates up to give at least the number of candidate events seen, and find that it is only $2 \times 10^{-6}$ (corresponding to 4.6 $\sigma$ if the errors on the backgrounds are Gaussian). Also shown in the table are measurements of the $t\bar{t}$ production cross section from each of the sets of cuts. The two values are consistent with each other.

Additional properties of the candidate events which indicate that the excess events are due to top quark pair production include the distribution of the 17 candidates between the 7 channels. The distribution is consistent with the top hypothesis at the 53% CL, which is very high. We calculated the cross section from each type of decay channel with a top quark mass of 200 GeV and found consistent results between each channel, which may be interpreted as an additional piece of evidence that the excess of events seen are due to $t\bar{t}$ pair production. These results are shown in Table 6, together with the cross section values calculated as a function of the top quark mass, using the results from the standard event selection cuts.

|  | Standard Cuts | Loose Cuts |
|---|---|---|
| Dilepton | 3 | 4 |
| Lepton+jets (shape) | 8 | 23 |
| Lepton+jets (*b* tag) | 6 | 6 |
| Total # of Candidates | 17 | 33 |
| MC $t\bar{t}$ Signal | $4.7 \pm 0.7$ | $6.3 \pm 0.9$ |
| Background | $3.8 \pm 0.6$ | $20.6 \pm 3.2$ |
| Probability | $2 \times 10^{-6}$ (4.6 $\sigma$) | 0.023 (2.0 $\sigma$) |
| $\sigma(t\bar{t} \to X)$ [pb] | $6.3 \pm 2.2$ | $4.5 \pm 2.5$ |

Table 5  Summary of search for $t\bar{t}$ production in all seven decay channels. The $t\bar{t}$ MC calculation is for a top quark of mass 200 GeV, as is the measured cross section.

Figure 3 shows the results of the DØ measurement of the top quark cross section as a function of the top quark mass, with a resummed next-to-leading-order (NLO) theoretical calculation of the cross section shown superposed for comparison[12].

The results from the DØ discovery of $t\bar{t}$ pair production were announced on March 2nd 1995 at Fermilab and appeared in Physical Review Letters on 3rd April 1995[13]. A similar publication from the CDF collaboration appeared simultaneously[14]. They measured the cross section as $6.8^{+3.6}_{-2.4}$ pb for a top quark mass of $176 \pm 8 \pm 10$ GeV.

| $t\bar{t}$ Decay Channel ($m_t$ = 200 GeV) | Cross Section [pb] | Top Quark Mass [GeV] | Cross Section [pb] |
|---|---|---|---|
| Dilepton | 7.6 ± 5.8 | 140 | 16.2 ± 5.6 |
| Lepton+jets (shape) | 4.9 ± 2.5 | 160 | 10.8 ± 3.7 |
| Lepton+jets ($b$ tag) | 8.9 ± 4.8 | 180 | 8.2 ± 2.9 |
| All channels | 6.3 ± 2.2 | 200 | 6.3 ± 2.2 |
|  |  | 220 | 5.1 ± 1.7 |
|  |  | 240 | 4.3 ± 1.5 |

Table 6   $t\bar{t}$ production cross section for each of the decay channels when $m_t$ = 200 GeV, and as a function of top quark mass from all channels combined.

## 5. What's New Since the Discovery of Top?

(a) *Overview of Work in Progress*

Since the discovery of the top quark by DØ in March 1995, DØ has approximately doubled its data set, from ~50 pb$^{-1}$ to ~100 pb$^{-1}$. Much work is going into analyzing this additional data and new results will be available shortly, but are beyond the scope of this paper. We have released preliminary results of a search for $t\bar{t}$ production in the alljets channel which will be discussed briefly here. We have also made a first preliminary measurement of the top quark mass using the dilepton events, which will be presented in a later section of this paper. New search techniques are in progress, including the use of neural networks and probability density estimators. And finally, we have begun a search for single top quark production[15].

(b) *Search in the Alljets Channel*

In the alljets channel, both $W$ bosons from top decay hadronically, leading to a final state with at least six jets ($b\bar{b}q\bar{q}'q\bar{q}'$) where the $q$ quarks are light. The advantage of searching in this channel is that it contains ~44% of the total branching fraction. The disadvantage is that there is almost overwhelming background from multijet events with at least six jets for which there is no Monte Carlo simulation. The cross section for multijet events with ≥6 jets is ~1000 × the $t\bar{t} \to$ alljets cross section. Event selection proceeded using topological techniques, coupled with $b$ tagging. The goal of the search was to obtain a signal-to-background ratio of 1:1, with a few signal events remaining.

To make the search, we required at least six jets in an event within the pseudorapidity region $|\eta| < 2$ and with $E_T > 15$ GeV, but we relaxed the transverse energy requirement on the sixth jet down to $E_T > 10$ GeV to maintain good efficiency. We then performed a grid search on several topological selection parameters such as aplanarity, centrality, the average number of jets in an event and the transverse scalar energy of all the jets except the two most energetic ones ($H_T^{3j}$).

Preliminary results from the alljets search are shown in Table 7 for two sets of cuts, the standard set designed to maximize the signal-to-background ratio, and a looser set designed to allow in some background so that it can be compared with the background predictions. The next step in this analysis is to calculate a cross section using the results from the search when two *b* tags are applied. When only one tag is found, then the $t\bar{t}$ cross section is measured to be $2.2 \pm 4.7$ pb using the standard cuts and $5.5 \pm 4.8$ pb from the looser cut search, for a top quark of mass 200 GeV. These values are consistent with each other, with the theoretical expectation and with the results from the searches in the dilepton and in the lepton+jets channels.

|  | Standard Cuts | Loose Cuts |
|---|---|---|
| no *b* tag |  |  |
|    MC $t\bar{t}$ Signal | 10.4 | 43.4 |
|    Data | 102 | 1483 |
| 1 *b* tag |  |  |
|    MC $t\bar{t}$ Signal | 2.3 | 9.5 |
|    Data | 4 | 50 |
|    Background | $3.2 \pm 0.3$ | $41.7 \pm 1.4$ |
|    Cross Section [pb] | $2.2 \pm 4.7$ | $5.5 \pm 4.8$ |
| 2 *b* tags |  |  |
|    MC $t\bar{t}$ Signal | 0.7 | 1.4 |
|    Data | 1 | 3 |
|    Background | 0.2 | 1.2 |

Table 7   Preliminary results from the search for $t\bar{t}$ production in the alljets channel. Predicted yields for Monte Carlo $t\bar{t}$ signal are for a top quark of mass 200 GeV.

## 6. Measuring the Top Quark Mass

(a) *Introduction*

We have used our candidate top events to measure the top quark mass using the decay kinematics. We have developed three techniques which I present here. The first uses the lepton+jets events from the time of the discovery and makes a fit using kinematic constraints. The second method uses the mass of the *W* boson reconstructed from the two jets into which it decayed as a calibration point for reconstructing the top quark mass from its decay products. This method is still in progress and has not yet lead to a new result on the top quark mass. The third technique uses the dilepton candidate events from a larger data set than that used for the discovery, and makes a fit to extract the top mass. This analysis leads to a new preliminary value for the top quark mass.

(b) *Mass Fitting in Lepton+Jets Events*

In lepton+jets events, the process is $p\bar{p} \to t\bar{t}X \to (bl\nu) + (\bar{b}q\bar{q}') + X$, so there are at least four jets in the final state, plus the lepton and some missing transverse energy from the passage of the neutrino out of the detector. The mass fitting technique is to make a two constraint (2C) kinematic fit, by smearing the energies of the jets and lepton (using measured energy resolutions) to force reconstruction of the two $W$ bosons with $m_{l\nu} = m_{q\bar{q}'} = m_W$ and then to force the top quark and $\bar{t}$ antiquark to have the same mass as each other $m_t = m_{\bar{t}}$. Sixteen variables are used in the fit: missing transverse energy, and $(E,\eta,\phi)$ for each of the five objects in the event $(b,\bar{b},q,\bar{q}',l)$. There is one unmeasured variable, the $z$ component (along the beamline) of the neutrino momentum. It is not possible to unambiguously reconstruct $p_z^\nu$, since the underlying event from the proton-antiproton interaction deposits much energy in $z$ which is not measured. For this fit, only the four highest $E_T$ jets with $|\eta| < 2.5$ are used. Attempting to combine additional jets with these highest energy ones, under the assumption that the extra jets might be final state radiation worsens the mass resolution. This is because although some of these jets are indeed final state radiation, they are sometimes recombined with the wrong jet, and also there is a significant amount of very high energy initial state radiation which should not be added in with the decay product jets from the tops. The jets were reconstructed using a narrow cone radius of 0.3, where the cone radius is defined as $R = \sqrt{\Delta\eta^2 + \Delta\phi^2}$. Corrections were applied to the jet energy for out of cone gluon radiation. The narrow cone was chosen to optimize the correspondence between reconstructed jet and initial state parton from Monte Carlo studies.

Figure 5 shows distributions of the fitted top quark mass with (a) exact parton energies from the Monte Carlo, and no initial or final state radiation, (b) a full GEANT[16] simulation of the DØ detector applied to the final state partons, thus smearing the energies, and occasionally merging or splitting the jets, (c) the additional detrimental effects of including final state radiation in the simulation and (d) the full simulation with both initial and final state radiation as well as energy smearing from a full detector simulation. In each plot, the shaded histogram shows the mass distribution when the correct jets are combined to reconstruct the $W$, $t$ and $\bar{t}$, and the black line histogram shows the mass distribution when the up to three combinations of jets with fit $\chi^2 < 7$ are chosen. It can be seen from Fig. 5 (a) that the intrinsic resolution on the top quark mass is rather good, at about 10 GeV, and the mean fitted mass is close to the top mass used in the simulation. Choosing the wrong combination of jets broadens this peak slightly. In plot 5 (b), the resolution of the DØ calorimeter widens the mass peak more, and choosing the wrong jet combinations significantly broadens the peak and shifts it to a lower mean mass. In 5 (c) the addition of final state radiation to the simulation shifts the peak to an even lower mass as energy is lost from the reconstruction and sometimes a gluon jet is more energetic than a quark jet from the $W$ decay, and so is mistakenly used in the fit. Finally, in 5 (d) the effects of initial state radiation are shown. Using an initial state gluon jet instead of one from a top decay can increase the reconstructed top quark mass, and so a high mass tail is added to the mass distribution.

The reconstructed or fitted mass for each top candidate event has to be converted into an actual top quark mass, by making a correction for the tendency for the mass peak to be shifted by initial and final state gluon radiation. By measuring the mean fitted top mass as a function of the input Monte Carlo top mass, one can make this correction. The relation used to make this correction is shown in Fig. 6.

The results of making kinematic mass fits to the lepton+jet event candidates is as follows. Of the 14 events which pass the standard cuts, 11 have at least four jets in the central pseudorapidity region and are successfully fitted. It is calculated that of these events, 2.1 ± 0.4 are actually from background sources. The true top quark mass using these events is found to be 199 GeV. Of the 29 events which pass the loose cuts, 24 have at least four central jets and are successfully fitted. There are 11.6 ± 2.2 background events in this sample. The true top quark mass is found to be 199 GeV, in excellent agreement with the result from the standard cuts analysis. The statistical and systematic errors are smaller using the loose cuts, and the final result is therefore taken from that analysis:

$$m_t = 199 \,^{+19}_{-21} \text{ (stat) } \,^{+14}_{-21} \text{ (syst) } = 199 \,^{+24}_{-30} \text{ GeV}$$

The systematic error is dominated by the 10% contribution from the uncertainty in the jet energy scale.

The upper two plots of Fig. 7 show in the shaded histograms the distribution of fitted masses of the candidate lepton+jets events for both the standard and loose event selection cuts. The outlined bins show which events have a muon tag (4 events which pass both the loose and standard cuts). The jet closest to the tagging muon was forced to be labeled as a *b* jet when combining jets in the mass fitting. The dotted curve shows the best fit to the distribution for the $t\bar{t} \to l + \text{jets}$ candidates and the dashed curve shows the best fit to the background. The solid curve is the sum of the other two curves and shows the best fit to the data. The position of the peak of the best fit curve to the signal does not change significantly if the number of background events is held fixed or is allowed to vary as a free parameter in the fit.

The lower two plots in Fig. 7 show the negative log likelihood of the fit to the signal as a function of the true top quark mass. The fits to the signal shown in the previous two plots were repeated over the full range of input top quark mass values, and the quality of the fit evaluated at each point (in 10 GeV intervals). The fitted masses were than converted to true masses using the relations shown in Fig. 6 and the fit quality plotted versus this true mass. The minima of the parabola-like curves shown here correspond to the best measurements of the top quark mass, both at 199 GeV.

(b) *An Aside – Properties of the $t\bar{t}$ Events*

Now that we have reconstructed the $t\bar{t}$ pairs from their decay products, we can study properties of the $t\bar{t}$ system such as the invariant mass of the pair and the mean transverse momentum of the top quarks. There are many theoretical extensions to the Standard Model which affect these distributions and so they will become of considerable interest in the future. At the moment there are too few events to make any precise differentiation between the various models. The upper two plots in Fig. 8 show the invariant mass distribution of the top candidates for the standard and loose cuts. The data is shown in the black crosses. The darker gray shaded histogram shows the predicted signal and the lighter gray histogram shows the predicted signal plus background. The data are consistent with the Standard Model prediction within errors.

The lower two plots in Fig. 8 show the mean transverse momentum of the reconstructed tops in these events for both standard and loose cuts. As before, data is represented by the crosses and Monte Carlo calculations of the signal and signal+background by the shaded histograms. No anomalies are seen.

(c) *($m_t$–$m_W$) Analysis*

The motivation for developing the following technique for extracting the mass of the top quark from the candidate events is to see if we can find a mass peak for the *W* boson from the two jets it decayed into. We will then be able to calibrate the top quark mass against the known *W* boson mass. This cannot be done using the 2C kinematic mass fitting method discussed previously, since explicit *W* mass constraints were applied. (The jet energies were adjusted until they reconstructed to give a perfect 80 GeV *W* boson.) The method discussed here will not constrain the *W* mass.

As before, only the four highest transverse energy jets are used in the reconstruction, this time using a cone of radius 0.5 to identify the jets. The wider cone was chosen so as not to be biased by the jet energy corrections, which had been tuned to reproduce the *W* boson mass. For each of the four permutations of how to combine the jets (two permutations if there was a *b* tag), we defined two top masses $m_t^{\text{lep}}$ and $m_t^{\text{had}}$ and a dijet mass $m_W$. We assigned a weight to each permutation based on the difference of top masses between the leptonic and hadronic sides of the decay,| $m_t^{\text{lep}}$ – $m_t^{\text{had}}$ |. We then plotted $\langle m_t \rangle$ versus $m_W$ for each weighted permutation of every event. We did not adjust the dijet energies and the dijet pairs were not selected for consistency with the *W* mass. The results of this procedure are shown in Fig. 9. The upper two plots show the ($m_t$–$m_W$) plane for (a) 200 GeV mass top HERWIG[17] Monte Carlo and (b) background *W*+jets VECBOS[11] MC events and multijets data. A clear peak at ($m_t$ = 190 GeV, $m_W$ = 77 GeV) as expected for the top MC shows that the method is successful at reconstructing both the top quark and the *W* boson. The nine highest bins in this region are lighter shaded in the figure to distinguish them. The peak in this first plot is projected onto the $m_t$ and $m_W$ axes in the lower two plots. No such peak is seen in the distribution from the backgrounds in the upper right-hand Lego plot, where the jets are mainly from lower energy gluons.

The data and best fit combination of backgrounds and $t\bar{t}$ signal with $m_t$ = 200 GeV are shown in the ($m_t$–$m_W$) plane in Fig. 10. The same nine bins are lightly shaded as in Fig. 9 where the $t\bar{t}$ MC signal peaked, to guide the eye. One can clearly see that the data selected with the loose cuts is a mixture of the $t\bar{t}$ signal and the backgrounds, with peaks in both regions of the plane.

So finally, we take the event distributions shown in Fig. 10 and project them onto the two axes to see if there is evidence for the decay of a *W* boson as indicated by a peak in the dijet mass distribution. We also look for a peak in the region of a heavy top quark on the other axis. These projections are shown in Fig. 11. On the upper plot, a cut has been made on events which have a dijet mass below 58 GeV, in order to remove background. A peak is seen in the reconstructed top mass distribution consistent with the top mass found from kinematic fitting. In the lower plot we make a cut on events with a reconstructed top quark mass less than 150 GeV to reject background. We now see a clean peak in the data above the predicted background, centered just below the *W* boson mass as expected.

This analysis technique has not yet been used to measure the top quark mass from the reconstructed dijet *W* mass. However, we can say that a $W \rightarrow$ jetjet peak is seen in the data, which is expected if the heavy particles decaying are top quarks. The probability of the background fluctuating up to give the observed peak in the $(m_t$–$m_W)$ plane is only 1.3%.

(d) *Dilepton Mass Analysis*

DØ has made a preliminary measurement of the top quark mass using the dilepton events. This technique is based on one proposed by Kondo[18] and later by Dalitz and Goldstein[19]. The method has been extended by DØ to use the missing transverse energy, and was published applied to one $t\bar{t} \rightarrow e\mu X$ event from the first collider run[2]. For the analysis presented here, a data sample of ~72 pb$^{-1}$ has been used. In this sample, 5 dilepton candidates have been found (one *ee*, two *eµ* and one *µµ* events) with an estimated background of one event. To make the mass fit, we assume the two *b* jets are the two highest transverse energy jets in the event. Monte Carlo ensembles of 5 event experiments are run to obtain the statistical error of the measurement. The systematic error as before comes mainly from the jet energy scale, and also from the choice of MC generator. The upper plot of Fig. 12 shows the best fit resolution functions for the data (solid curve) and the background (dashed curve), as well as the fitted mass for the five candidate events. The fit is repeated for many possible top quark masses and the log likelihood of the fits is shown in the lower plot. The maximum log likelihood occurs for a true top mass of 145 GeV, where the best fitted mass of 162 GeV has been corrected to give the true mass in a similar manner to that shown in Fig. 6. The preliminary new result from mass fitting the five dilepton event candidates is:

$$m_t \approx 145 \pm 25 \text{ (stat)} \pm 20 \text{ (syst)} \approx 145 \pm 32 \text{ GeV}$$

If one uses only the two *eµ* events, which have the smallest chance of being background, then the result is almost exactly the same as when all five dilepton candidates are used.

## 7. The Future of Top Physics at DØ

Top physics is being actively pursued at DØ in many directions. We are extending the analyses from the 50 pb$^{-1}$ of data presented here to use the ~100 pb$^{-1}$ which we accumulated by July 1995. Results from this large data set should be forthcoming in January 1996. In addition to more data, we are also expanding our analysis techniques to gain higher efficiency for event selection and better background rejection. We are improving our lepton identification methods, and extending the *b* tagging method to encompass electrons as well as muons. Much work is also going into improving the mass measurement techniques. A search for electroweak single top production has started. Finally, we are actively preparing for the next run in 1999, which should bring us 2 fb$^{-1}$ of data, together with a significantly upgraded detector. We will have a central 2 T magnetic field, and a silicon tracker vertex detector which will enable us to do detached secondary vertex tagging of B hadron decays in jets, as well as the semileptonic tagging we currently use. The prospects for top physics in the next run are very exciting.

## 8. Summary of Top Physics at DØ

The DØ collaboration discovered the top quark at the Fermilab Tevatron collider in March 1995. In a ~50 pb$^{-1}$ data set, we found 17 candidate $t\bar{t}$ events with a background of $3.8 \pm 0.6$ events using our standard event selection cuts, and 33 events with a background of $20.6 \pm 3.2$ events using a loose set of cuts. We measured the top quark mass using two separate techniques. From the lepton+jets events, we found $m_t = 199^{+24}_{-30}$ GeV, and from the dilepton events we made a preliminary measurement of $m_t \approx 145 \pm 32$ GeV. At the top mass of 199 GeV, we measured the $p\bar{p} \to t\bar{t}X$ production cross section to be $6.4 \pm 2.2$ pb.

## Acknowledgements

I would like very much to thank the people from Moscow State University Nuclear Physics Institute for inviting me to this workshop. My visit to Russia was extremely interesting and enjoyable, due to the kindness and consideration of the members of the organizing committee to whom I am very grateful.

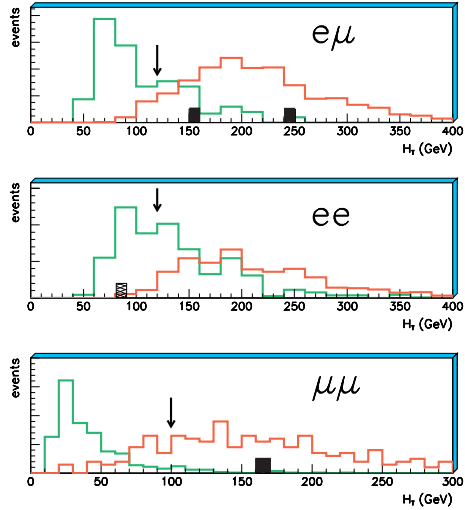

Fig. 1  Dilepton $H_T$ distributions. The dark gray line at high $H_T$ shows the Monte Carlo $t\bar{t}$ distribution ($m_t = 180$ GeV), the light gray line shows the background. The arrows indicate the $H_T$ cut value for each channel. The black rectangles show the candidate events and the hatched rectangle shows an $ee$ event which fails only the $H_T$ cut.

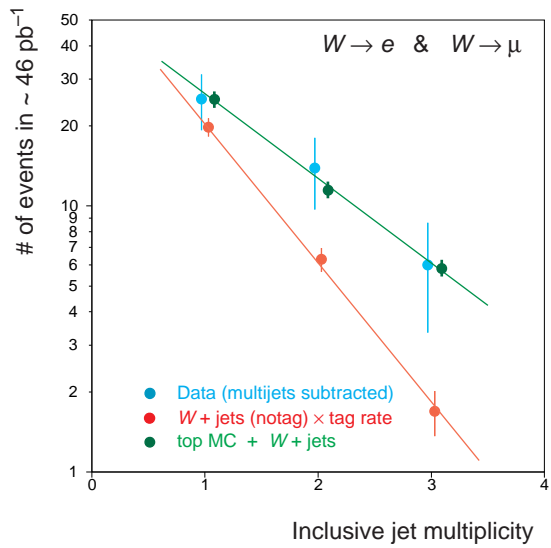

Fig. 2  Lepton+jets/muon tag events versus the inclusive jet multiplicity, before the $H_T$ cut was applied. The pale gray points with large error bars show the data (with multijets subtracted). The mid-gray points show $W$+jets data where there is no tagging muon found, convoluted with the measured tagging rate. The dark gray points show how the $W$+jets/no tag × tag rate data become consistent with the total tagged data when $t\bar{t}$ Monte Carlo events are added.

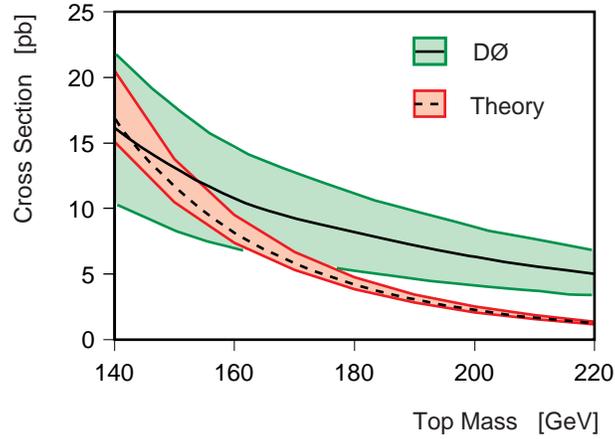

Fig. 3  $t\bar{t}$ production cross section as a function of top quark mass. The solid line is the DØ measurement and the dashed line is a resummed NLO calculation from theory.

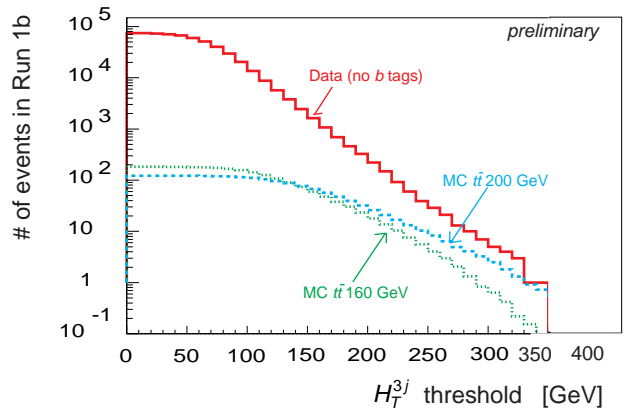

Fig. 4  Distribution of events with at least six central jets in the variable $H_T^{3j}$ (defined in text). The dark solid line shows the Run 1b data (~25 pb$^{-1}$), the dashed line shows the Monte Carlo $t\bar{t}$ distribution for $m_t = 200$ GeV and the dotted line shows the distribution for $m_t = 160$ GeV. At large $H_T^{3j}$, the signal-to-background ratio starts to improve significantly.

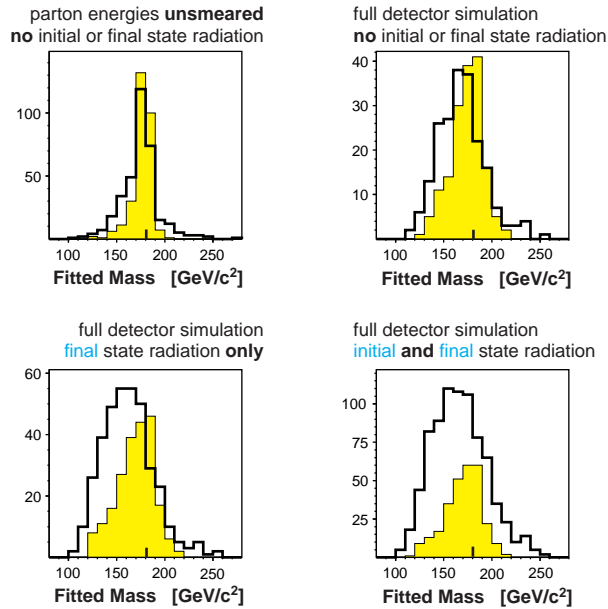

Fig. 5　Distribution of fitted top quark mass from 180 GeV $t\bar{t} \rightarrow e +$ jets events from an ISAJET simulation, with loose cuts applied. The shaded histograms show the distributions using the correct jet combinations. The black line histograms show up to three low $\chi^2$ solutions trying all possible jet combinations.

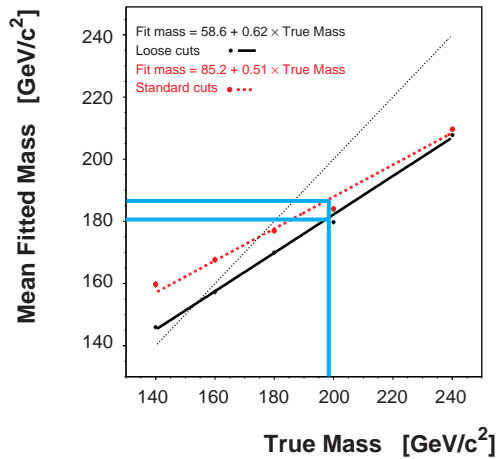

Fig. 6　Mean fitted top quark mass versus true top quark mass. The solid black line is for the loose cuts and the dashed line is for the standard cuts. The standard cuts gain extra background rejection in the low mass region, but also throw out more low mass top events, thus biasing the distribution. The thin dotted line is to guide the eye for a one-to-one relation. The gray lines from mean fitted masses of 182 GeV (loose cuts) and 187 GeV (tight cuts) show how both measurements give a true top mass of 199 GeV.

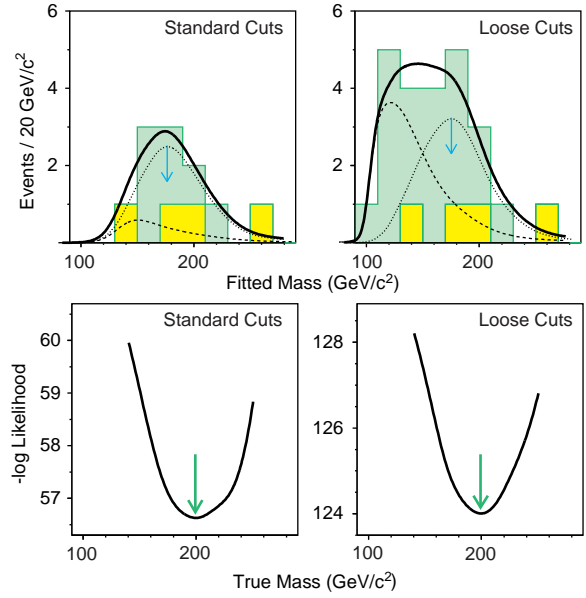

Fig. 7 The upper plots show for standard and loose cuts the fitted top quark mass for the data (shaded histogram) and the best fits to the signal (dotted curve), background (dashed curve) and total (solid curve). The outlined histogram bins are four *b* tagged events. The lower plots show the negative log likelihood of the true mass for the standard and loose cuts, with minima at 199 GeV on both plots. The width of the parabola-like curve gives the statistical error of the measurement.

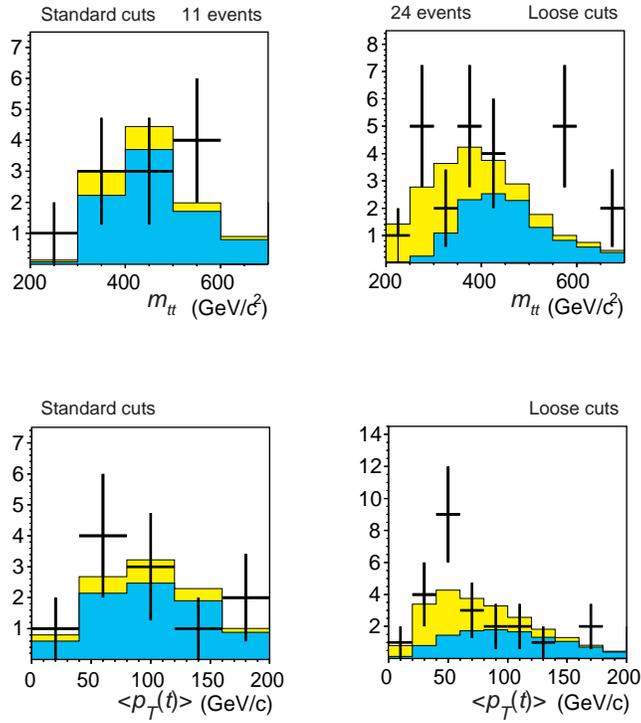

Fig. 8   Upper plots show the invariant mass of the $t\bar{t}$ system with data (crosses), Monte Carlo signal, 200 GeV ISAJET simulation (dark histogram) and MC signal+background (light histogram). Lower plots show the mean transverse momentum of the top quarks in these candidate events.

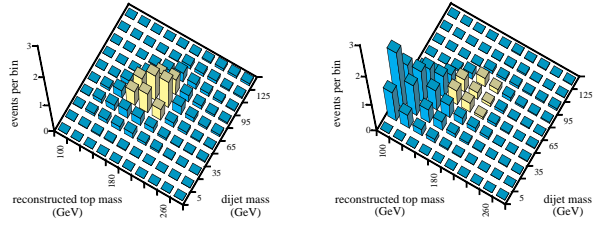

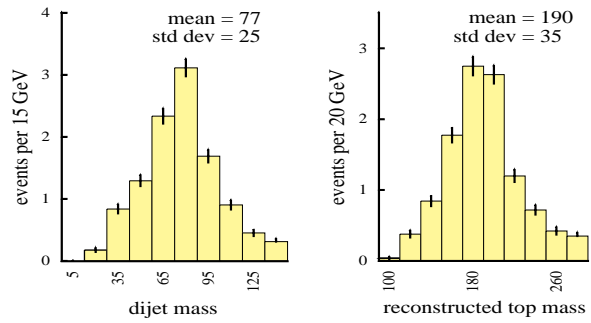

Fig. 9  Upper plots show the distribution of Monte Carlo signal (left plot) and background (right plot) in the ($m_t$–$m_W$) plane. Lower plots show the projections of the top MC signal onto the $m_W$ axis and the $m_t$ axis.

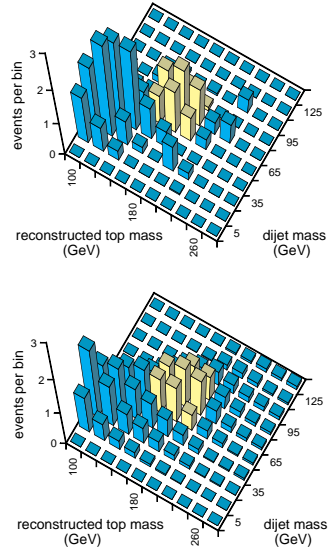

Fig. 10  Data with loose cuts (upper plot) and best fit backgrounds plus 200 GeV HERWIG MC $t\bar{t}$ signal (lower plot) in the ($m_t$–$m_W$) plane. A clear peak can be seen in the data away from the backgrounds.

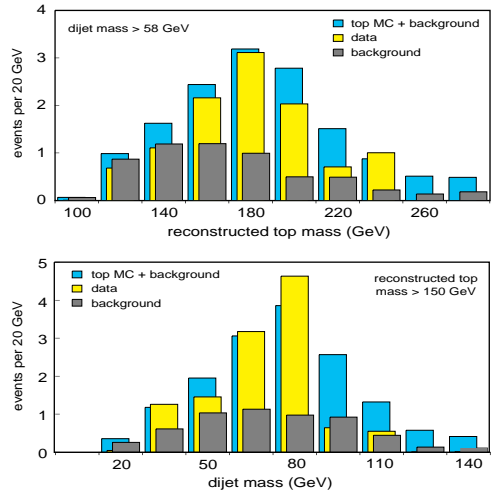

Fig. 11  Projections of the $t\bar{t} \rightarrow l +$ jets data onto the reconstructed top mass axis (upper plot) and onto the dijet mass axis (lower plot) of the ($m_t$–$m_W$) plane, with cuts applied to reject background. The dark gray histograms at the front are background, the light gray histograms are data and the mid gray histograms at the rear are 200 GeV HERWIG $t\bar{t}$ MC + background.

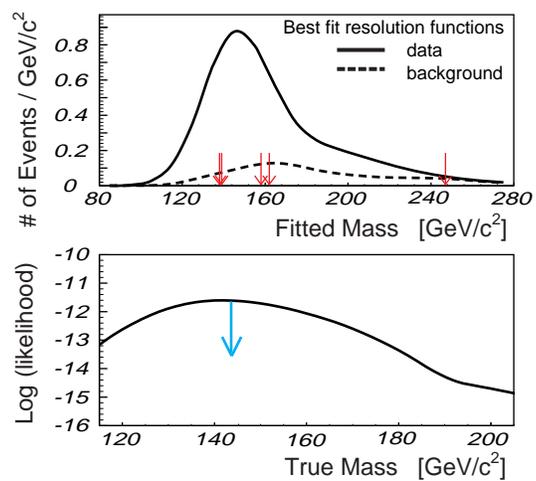

Fig. 12　The upper plot shows distributions of fitted mass for the data (solid line), backgrounds (dashed line) and candidate events (arrows). The lower plot shows the log likelihood distribution of the fits, with the peak at a true top quark mass of 145 GeV.